# GARNET RING MEASUREMENTS FOR THE FERMILAB BOOSTER 2$^{ND}$ HARMONIC CAVITY*

J. Kuharik [†], J. Dey, K. Duel, R. Madrak, A. Makarov, W. Pellico, J. Reid, G. Romanov,
M. Slabaugh, D. Sun, C. Y. Tan, I. Terechkine
Fermilab, Batavia, IL, 60510, USA

*Abstract*

A perpendicularly biased tuneable 2nd harmonic cavity is being constructed for use in the Fermilab Booster. The cavity's tuner uses National Magnetics AL800 garnet as the tuning media. For quality control, the magnetic properties of the material and the uniformity of the properties within the tuner must be assessed. We describe two tests which are performed on the rings and on their corresponding witness samples.

## INTRODUCTION

As part of Fermilab's Proton Improvement Plan (PIP), a perpendicularly biased 2$^{nd}$ harmonic cavity is being constructed to help minimize beam losses. This cavity is fundamentally different from most other ferrite tuned cavities at Fermilab, in that its bias magnetic field is perpendicular to the RF magnetic field.

The tuner for the cavity, which is discussed in [1], contains five garnet rings. Each of the garnet rings is assembled on a 3mm thick alumina substrate ring by epoxying together (and to the substrate) eight sectors of AL800 garnet. Magnetic properties of the material in each sector are measured using witness samples that accompany each sector.

To qualify each ring assembly a special setup was designed and procured. The setup consists of an RF cavity and a magnetic bias system. This allows the extraction of the average values of the magnetic properties of the material, projected tuning range, and the expected power loss in the 2$^{nd}$ harmonic cavity.

## GARNET RINGS

Each of the five rings in the tuner section of the cavity uses National Magnetics AL800 material, which is aluminium doped garnet. The saturation magnetization ($4\pi M_s$) is 800 G and the Curie temperature is 210 °C. The OD and ID of the rings are 13.386" and 8.268", respectively, and the thickness is 0.827". The garnet rings are made from eight sectors which are epoxied together because the vendor's oven could not accommodate a full ring. To simplify heat removal from the garnet material, each ring is epoxied to a 0.118" thick alumina (99.5%) ring using Stycast 2850FT, Catalyst 9. Each garnet sector is cut from a brick along with a small "witness sample", to be used for quality control. A drawing and photograph of one tuner ring assembly is shown in Fig. 1.

## WITNESS SAMPLE TESTING

As each garnet ring is made using eight sectors, a certain degree of confidence is needed that the values of the magnetic properties in each sector are close, if not identical. The static permeability was measured for each witness sample using a specially designed setup. A sketch and photographs are shown in Fig. 2. The sample measurement circuit is shown in Fig. 3. Two coils are used in the setup: the excitation coil and the signal coil. The magnetic field strength $H$ is calculated knowing the properties of the excitation coil and the current. The voltage induced in the signal coil during the current ramp determines the magnetic flux through the witness sample. From these two quantities, and dimensions of the test setup, the permeability is extracted. Details of the test setup and measurement circuit are discussed in greater detail in [2] and [3].

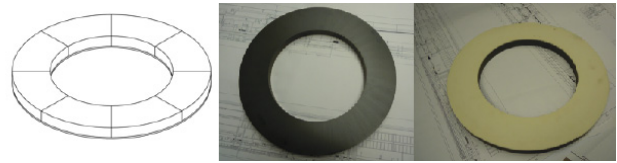

Figure 1: Drawing and photographs of the garnet and alumina sides of a tuner ring. The ring OD is 13.386".

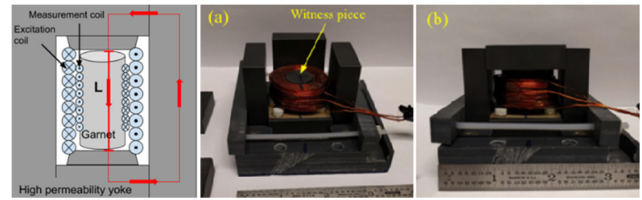

Figure 2: Sketch and photographs of the witness sample test setup.

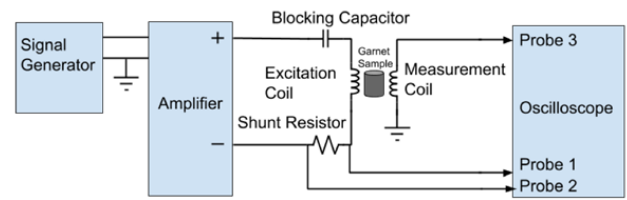

Figure 3: Circuit used in the witness sample permeability measurements.

Measurements have been performed for 83 witness samples and the permeability is satisfactorily uniform. This consistency gives confidence that the sectors of the fully assembled garnet rings do not have significant variation.

## TUNER RING TESTING

The test setup for the ring measurement was designed to ensure that the magnetic field in the garnet was as uniform as possible. The specially designed RF cavity and bias system are shown in Fig. 4. Two weakly coupled probes are used to measure frequency and $Q$. These vary depending on the bias magnetic field in the ring, which is generated by the solenoid.

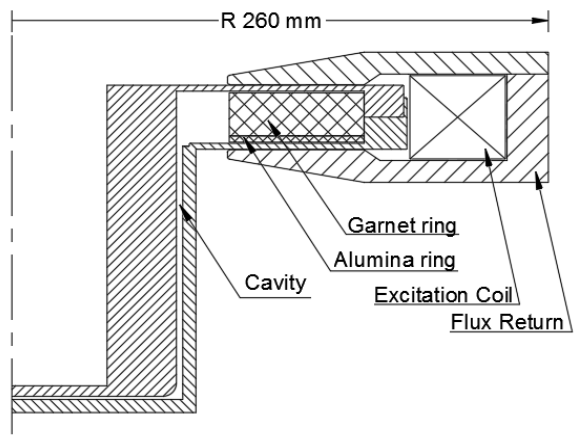

Figure 4: Schematic of the tuner ring test cavity with magnet.

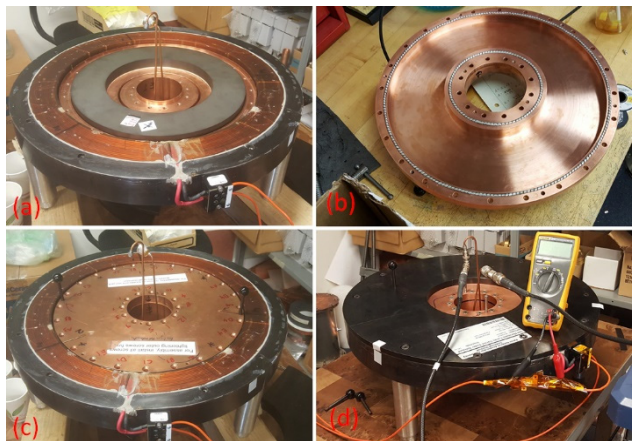

Figure 5: Photographs of the tuner ring test cavity (a) with tuner ring and shorting lid off, (b) inside of shorting lid, (c) with shorting lid on, (c) fully assembled with flux return top in place.

For measurement, the garnet rings are placed in the large OD section of the cavity at the shorted end. On the OD, the shorting contact is made by a 0.003" interference fit between the convex edge of the cavity top lid and the lip on the outer shell. The contact is maintained by twenty ¼-20 screws. In addition, a tin-plated beryllium copper gasket (Spira-Shield®) is used in grooves between the faces of the lid and the outer and inner conductors (not shown in Fig. 4). On the inner conductor, the contact is made between the faces of the lid and the center conductor, with twelve ¼-20 screws.

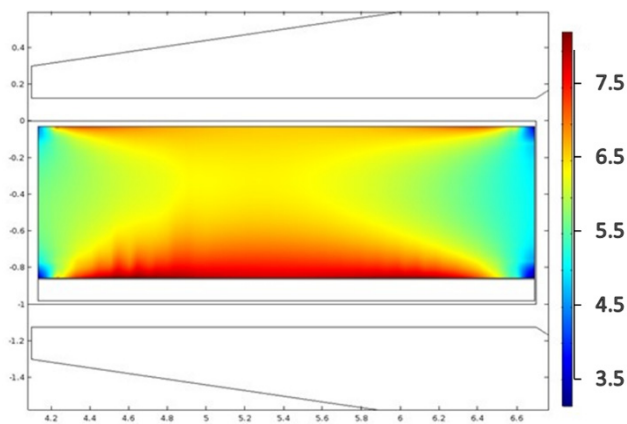

Figure 6: Modelling results for permeability in the garnet when the solenoid bias is ~1000 A-turns and the cavity resonant frequency is ~ 70 MHz.

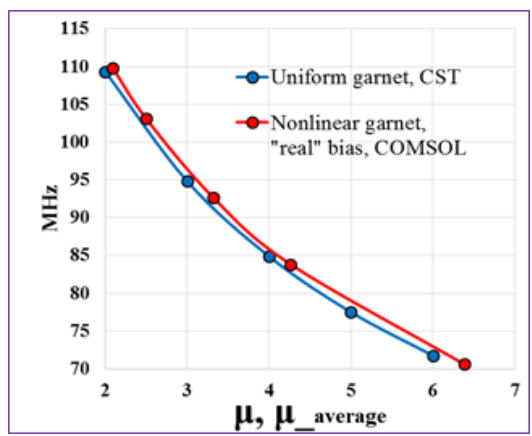

Figure 7: Modelling results for cavity frequency as a function of the garnet permeability. In one case (CST), it was assumed that the permeability in the garnet was uniform. In another case (COMSOL) it was not uniform.

The biasing solenoid for the test cavity is made from a rectangular cross section round coil with 224 turns of 10-gauge square copper wire. The flux return is made from 1010 low carbon steel. Photographs are shown in Fig. 5. Fig. 6 shows the model prediction of the variation in the permeability of a tuner ring when the test cavity is at low bias. Fig. 7 shows the model prediction for the average permeability in the garnet as a function of frequency. The model assumed a permeability curve and losses which we had previously measured using smaller samples of garnet [4].

### Results

Figs. 8 and 9 show the measured resonant frequency and $Q$ of the test cavity with each of the five tuner rings. Two measurements are shown for each ring to quantify the repeatability of the measurements. Between successive measurements of the same ring, the cavity top was removed and then reinstalled. The results for the frequency agree well with predictions from simulation and show acceptable scatter, in that the differences between the measured values for one ring are similar to the differences for two different rings. The $Q$ measurements are also uniform from ring to ring, but in this case, it was not possible to match data with simulation. This may be because the loss coefficient $\alpha$ is not constant, although measurements presented in [4] seemed to indicate otherwise. Nevertheless, the rings are very uniform in their material properties indicating good quality control at the vendor.

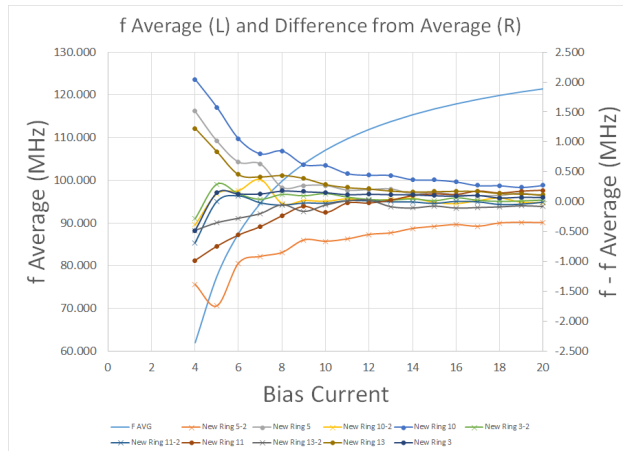

Figure 8: Measured frequency (average and difference from average) for the five tuner rings. The plot shows two measurements for each ring.

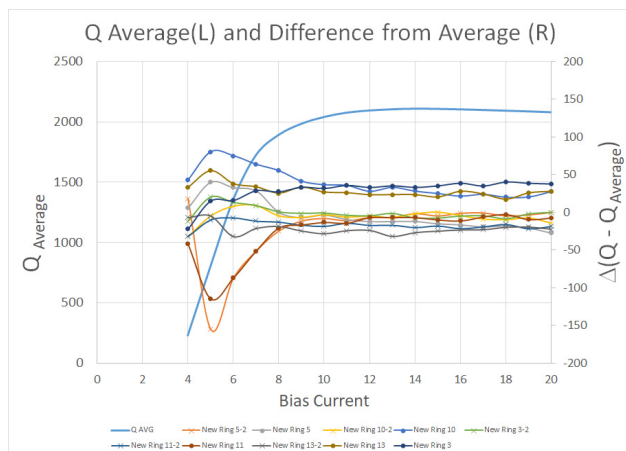

Figure 9: Measured $Q$ (average and difference from average) for the five tuner rings. The plot shows two measurements for each ring.

## CONCLUSION

We are in the process of constructing a 2nd harmonic cavity for the Fermilab Booster. We have measured the magnetic properties of the fully assembled tuner rings and their corresponding witness samples. The material is very uniform, and results agree well with simulation for the real part of the permeability. The losses in the garnet are also uniform, but the measured values are not completely understood.

The cavity will be tested early this summer and installed into the Booster during a planned shutdown. If successful, this will be the first operational broadband perpendicularly biased cavity and it will be a significant technical achievement for accelerators.

## ACKNOWLEDGMENTS

Many thanks to National Magnetics for the manufacture of our garnet and the assembly of the tuner rings with our required consistency and precision.